\documentclass[12pt]{article}
\usepackage{amstex}
\usepackage{amssymb}
\newcommand{\TITLE}[4]{\begin{center}
{\Huge \bf #1}\\[3ex]
{\Large \it #2} \\[1ex]
{\it #3}\\[4ex]
\parbox[t]{0.8\textwidth}{
\centerline{\bf Abstract}
\normalsize \it
\noindent
#4\vspace{4ex}
}
\end{center}}
\allowdisplaybreaks
\newcommand{\bs}{\boldsymbol}
\newcommand{\C}{\mathbb{C}}
\newcommand{\R}{\mathbb{R}}
\newcommand{\FS}{\mathbb{FS}}
\newcommand{\E}{\mathbf{E}}
\newcommand{\SL}{\mathop{\text{SL}}\nolimits}
\newcommand{\Herm}{\mathop{\text{Herm}}\nolimits}
\newcommand{\contr}{\mathop{\text{contraction}}\nolimits}
\newcommand{\tr}{\mathop{\text{trace}}\nolimits}
\begin{document}\large
\TITLE{Finslerian $\bs{N}$-spinor algebra}
{A.V. Solov'yov}
{Moscow State University}
{New mathematical objects called Finslerian $N$-spinors are discussed. The
Finslerian $N$-spinor algebra is developed. It is found that Finslerian
$N$-spinors are associated with an $N^2$-dimensional flat Finslerian space. A
generalization of the epimorphism $\SL(2,\C)\to\text{O}^\uparrow_+(1,3)$ to a
case of the group $\SL(N,\C)$ is constructed.}

\section*{Introduction}

Spinors as geometrical objects were discovered by \'E.~Cartan~\cite{1} in 1913.
One decade later, W.~Pauli~\cite{2} and P.A.M.~Dirac~\cite{3} rediscovered
spinors in connection with the problem of describing the spin of an electron.
From that time, spinors are intensively used in mathematics and physics.

In the classical works~\cite{4,5}, the concept of Cartan's 2-spinor was
generalized and the theory of spinors in an arbitrary $n$-dimensional
pseudo-Euclidean space was constructed. In this report, another generalization
of 2-spinors is proposed which leads to the Finslerian geometry. Originally,
such a generalization appeared within the so-called {\it relational theory of
space-time\/}~\cite{6,7}. However, the corresponding mathematical scheme has
also an independent meaning and will be presented below.

\section*{General formalism}

Let $\FS^N$ be a vector space of $N>1$ dimensions over $\C$ and
$$
[\cdot,\cdot,\dots,\cdot]\colon\underbrace{\FS^N\times\FS^N\times\cdots\times
\FS^N}_{\text{$N$ multiplicands}}\to\C\eqno(1)
$$
be a nonzero antisymmetric $N$-linear functional on $\FS^N$. The latter
means:\\
\indent
\phantom{ii}(i) there exist $\bs{\xi}_0$, $\bs{\eta}_0$, \dots, $\bs{\lambda}_0
\in\FS^N$ such that
$$
[\bs{\xi}_0,\bs{\eta}_0,\dots,\bs{\lambda}_0]=z_0\ne 0;\eqno(2)
$$
\indent
\phantom{i}(ii) for any $\bs{\xi}_1$, $\bs{\xi}_2$, \dots, $\bs{\xi}_N\in
\FS^N$,
$$
[\bs{\xi}_a,\bs{\xi}_b,\dots,\bs{\xi}_c]=\varepsilon_{ab\dots c}\,[\bs{\xi}_1,
\bs{\xi}_2,\dots,\bs{\xi}_N],
$$
where $a$, $b$, \dots, $c=1$, 2, \dots, $N$ and $\varepsilon_{ab\dots c}$ is
the $N$-dimensional Levi-Civita symbol with the ordinary normalization
$\varepsilon_{12\dots N}=1$;\\
\indent
(iii) for any $\bs{\xi}_1$, $\bs{\eta}_1$, $\bs{\xi}_2$, $\bs{\eta}_2$, \dots,
$\bs{\xi}_N$, $\bs{\eta}_N\in\FS^N$ and $z\in\C\,$,
\begin{align*}
[\bs{\xi}_1,\dots,\bs{\xi}_a+\bs{\eta}_a,\dots,\bs{\xi}_N]
&=
[\bs{\xi}_1,\dots,\bs{\xi}_a,\dots,\bs{\xi}_N]+[\bs{\xi}_1,\dots,\bs{\eta}_a,
\dots,\bs{\xi}_N],\\
[\bs{\xi}_1,\dots,z\bs{\xi}_a,\dots,\bs{\xi}_N]
&=
z[\bs{\xi}_1,\dots,\bs{\xi}_a,\dots,\bs{\xi}_N],
\end{align*}
where $a$ takes the values 1, 2, \dots, $N$.

We shall use the following terminology. The space $\FS^N$ equipped with
the functional (1) having the properties (i), (ii), and (iii) is called the
{\it space of Finslerian $N$-spinors}. The complex number $[\bs{\xi},
\bs{\eta},\dots,\bs{\lambda}]$ is respectively called the {\it scalar
$N$-product\/} of the Finslerian $N$-spinors $\bs{\xi}$, $\bs{\eta}$, \dots,
$\bs{\lambda}\in\FS^N$. 

It should be noted that $\bs{\xi}_0$, $\bs{\eta}_0$, \dots, $\bs{\lambda}_0$
are linearly independent. Indeed, if those were linearly dependent, one of the
Finslerian $N$-spinors $\bs{\xi}_0$, $\bs{\eta}_0$, \dots, $\bs{\lambda}_0$
would be a linear combination of the others and, in accordance with
(ii)--(iii), the scalar $N$-product $[\bs{\xi}_0,\bs{\eta}_0,\dots,
\bs{\lambda}_0]$ would be equal to zero. However, this is in contradiction with
(2). Thus, $\bs{\xi}_0$, $\bs{\eta}_0$, \dots, $\bs{\lambda}_0$ are linearly
independent, i.e., form a basis in $\FS^N$.

Let us introduce the notation $\bs{\epsilon}_1=\bs{\xi}_0$, $\bs{\epsilon}_2=
\bs{\eta}_0$, \dots, $\bs{\epsilon}_N=\bs{\lambda}_0/z_0$. It is evident that
the set $\{\bs{\epsilon}_1,\bs{\epsilon}_2,\dots,\bs{\epsilon}_N\}$ is a basis
in $\FS^N$. Due to (2) and (iii), its elements satisfy the condition
$$
[\bs{\epsilon}_1,\bs{\epsilon}_2,\dots,\bs{\epsilon}_N]=1.\eqno(3)
$$
We shall call such a basis {\it canonical}.

Let $\bs{\epsilon}_1^\prime$, $\bs{\epsilon}_2^\prime$, \dots,
$\bs{\epsilon}_N^\prime$ be arbitrary Finslerian $N$-spinors and
$$
\bs{\epsilon}_a^\prime=c_a^b\bs{\epsilon}_b\eqno(4)
$$
be their expansions into the canonical basis $\{\bs{\epsilon}_1,
\bs{\epsilon}_2,\dots,\bs{\epsilon}_N\}$; here $a$, $b=1$, 2, \dots, $N$,
$c_a^b\in\C\,$, and the summation is taken over the repeating index $b$.
With the help of (ii), (iii), (3), and (4), we find
$$
[\bs{\epsilon}_1^\prime,\bs{\epsilon}_2^\prime,\dots,\bs{\epsilon}_N^\prime]=
\det\text{C}_N,\eqno(5)
$$
where $\text{C}_N=\|c_a^b\|$. Since linear (in)dependence of
$\bs{\epsilon}_1^\prime$, $\bs{\epsilon}_2^\prime$, \dots,
$\bs{\epsilon}_N^\prime$ is equivalent to that of columns of the complex
$N{\times}N$ matrix $\text{C}_N$, the set $\{\bs{\epsilon}_1^\prime,
\bs{\epsilon}_2^\prime,\dots,\bs{\epsilon}_N^\prime\}$ is a basis in
$\FS^N$ if and only if $\det\text{C}_N\ne 0$. Moreover, it follows from
(5) that $\{\bs{\epsilon}_1^\prime,\bs{\epsilon}_2^\prime,\dots,
\bs{\epsilon}_N^\prime\}$ is a canonical one when $\det\text{C}_N=1$. Thus, if
$\text{C}_N$ runs the group $\SL(N,\C)$ of unimodular complex
$N{\times}N$ matrices, then $\{\bs{\epsilon}_1^\prime,\bs{\epsilon}_2^\prime,
\dots,\bs{\epsilon}_N^\prime\}$ runs the set $\E(\FS^N)$ of
canonical bases in $\FS^N$.

Let us express the scalar $N$-product of Finslerian $N$-spinors in terms of
their components with respect to {\it any\/} canonical basis $\{\bs{\epsilon}_1
,\dots,\bs{\epsilon}_N\}\in\E(\FS^N)$. By using (ii), (iii), (3), and the
expansions $\bs{\xi}=\xi^a\bs{\epsilon}_a$, $\bs{\eta}=\eta^b\bs{\epsilon}_b$,
\dots, $\bs{\lambda}=\lambda^c\bs{\epsilon}_c$, it is possible to show that
$$
[\bs{\xi},\bs{\eta},\dots,\bs{\lambda}]=\varepsilon_{ab\dots c}\,\xi^a\eta^b
\cdots\lambda^c,\eqno(6)
$$
where $\bs{\xi}$, $\bs{\eta}$, \dots, $\bs{\lambda}\in\FS^N$, $\xi^a$,
$\eta^b$, \dots, $\lambda^c\in\C\,$, $a$, $b$, \dots, $c=1$, 2, \dots,
$N$. In (6) as well as in the following formulas of this article, the summation
is taken over all the repeating indices. It is clear that the scalar
$N$-product (6) is zero if and only if $\bs{\xi}$, $\bs{\eta}$, \dots,
$\bs{\lambda}$ are linearly dependent Finslerian $N$-spinors.

Let us consider a mapping
\begin{align}
\bs{S}\colon\E(\FS^N)&\to\C\,^{N^{k+l+m+n}},\nonumber\\
\{\bs{\epsilon}_1,\dots,\bs{\epsilon}_N\}&\mapsto\bs{S}\{\bs{\epsilon}_1,\dots,
\bs{\epsilon}_N\}=\left(S_{a_1\!\dots a_m\dot{d}_1\!\dots\dot{d}_n}^{b_1\!\dots
b_k\dot{c}_1\!\dots\dot{c}_l}\!\{\bs{\epsilon}_1,\dots,\bs{\epsilon}_N\}\right)
\tag{7}
\end{align}
such that
\begin{align}
S_{a_1\!\dots a_m\dot{d}_1\!\dots\dot{d}_n}^{b_1\!\dots b_k\dot{c}_1\!\dots
\dot{c}_l}\!\{\bs{\epsilon}_1^\prime,\dots,\bs{\epsilon}_N^\prime\}
&=
c_{a_1}^{e_1}\!\cdots c_{a_m}^{e_m}
\overline{c_{\dot{d}_1}^{\dot{h}_1}}\!\cdots
\overline{c_{\dot{d}_n}^{\dot{h}_n}}
d^{b_1}_{f_1}\!\cdots d^{b_k}_{f_k}
\overline{d^{\dot{c}_1}_{\dot{g}_1}}\!\cdots
\overline{d^{\dot{c}_l}_{\dot{g}_l}}\nonumber\\
&\times
S_{e_1\!\dots e_m\dot{h}_1\!\dots\dot{h}_n}^{f_1\!\dots f_k\dot{g}_1\!\dots
\dot{g}_l}\!\{\bs{\epsilon}_1,\dots,\bs{\epsilon}_N\}\tag{8}
\end{align}
for any two canonical bases $\{\bs{\epsilon}_1,\dots,\bs{\epsilon}_N\}$,
$\{\bs{\epsilon}_1^\prime,\dots,\bs{\epsilon}_N^\prime\}\in\E(\FS^N)$ whose
elements are connected by the relations (4). Here all the indices (both
ordinary and dotted) run independently from 1 to $N$, the over-lines denote
complex conjugating, $d_b^a$ are the complex numbers satisfying the conditions
$c_a^b d_c^a=\delta^b_c$ ($\delta^b_c$ is the Kronecker symbol), $\det\|c^a_b\|
=\det\|d^a_b\|=1$, and $k$, $l$, $m$, $n$ are nonnegative integers.

Every mapping (7), which possesses the property (8), is called a {\it
Finslerian $N$-spintensor of a valency\/} $\left[{k\atop m}{l\atop n}\right]$.
The addition and multiplication of such $N$-spintensors are defined in the
standard way: if $\bs{S}$ and $\bs{T}$ have the valency $\left[{k\atop m}{l
\atop n}\right]$ while $\bs{U}$ has the valency $\left[{p\atop r}{q\atop s}
\right]$, then
\begin{align*}
(S+T)_{a_1\!\dots a_m\dot{d}_1\!\dots\dot{d}_n}^{b_1\!\dots b_k\dot{c}_1\!\dots
\dot{c}_l}\!\{\bs{\epsilon}_1,\dots,\bs{\epsilon}_N\}
&=
S_{a_1\!\dots a_m\dot{d}_1\!\dots\dot{d}_n}^{b_1\!\dots b_k\dot{c}_1\!\dots
\dot{c}_l}\!\{\bs{\epsilon}_1,\dots,\bs{\epsilon}_N\}\\
&+
T_{a_1\!\dots a_m\dot{d}_1\!\dots\dot{d}_n}^{b_1\!\dots b_k\dot{c}_1\!\dots
\dot{c}_l}\!\{\bs{\epsilon}_1,\dots,\bs{\epsilon}_N\}
\end{align*}
are the components of the sum $\bs{S}+\bs{T}$ while
\begin{align*}
(S\otimes U)_{a_1\!\dots a_{m+r}\dot{d}_1\!\dots\dot{d}_{n+s}}^{b_1\!\dots
b_{k+p}\dot{c}_1\!\dots\dot{c}_{l+q}}\!\{\bs{\epsilon}_1,\dots,\bs{\epsilon}_N
\}
&=
S_{a_1\!\dots a_m\dot{d}_1\!\dots\dot{d}_n}^{b_1\!\dots b_k\dot{c}_1\!\dots
\dot{c}_l}\!\{\bs{\epsilon}_1,\dots,\bs{\epsilon}_N\}\\
&\times
U_{a_{m+1}\!\dots a_{m+r}\dot{d}_{n+1}\!\dots\dot{d}_{n+s}}^{b_{k+1}\!\dots
b_{k+p}\dot{c}_{l+1}\!\dots\dot{c}_{l+q}}\!\{\bs{\epsilon}_1,\dots,
\bs{\epsilon}_N\}
\end{align*}
are those of the product $\bs{S}\otimes\bs{U}$ with respect to an arbitrary
canonical basis $\{\bs{\epsilon}_1,\dots,\bs{\epsilon}_N\}\in\E(\FS^N)$.
Notice that all Finslerian $N$-spintensors of the valency $\left[{k\atop m}
{l\atop n}\right]$ form an $N^{k+l+m+n}$-dimensional vector space over $\C\,$.

Let $\Herm(N)$ be an {\it$N^2$-dimensional vector space over\/}
$\R$ consisting of Finslerian $N$-spintensors $\bs{X}$ of the valency
$\left[{1\atop 0}{1\atop 0}\right]$ whose components satisfy the Hermitian
symmetry conditions
$$
X^{b\dot c}\{\bs{\epsilon}_1,\dots,\bs{\epsilon}_N\}=\overline{X^{c\dot b}
\{\bs{\epsilon}_1,\dots,\bs{\epsilon}_N\}}\eqno(9)
$$
for any $\{\bs{\epsilon}_1,\dots,\bs{\epsilon}_N\}\in\E(\FS^N)$. Besides, let
$\{\bs{E}_0,\bs{E}_1,\dots,\bs{E}_{N^2-1}\}$ be a basis in $\Herm(N)$ and $\{
\bs{\epsilon}_1,\bs{\epsilon}_2,\dots,\bs{\epsilon}_N\}$ be a canonical one in
$\FS^N$. With each $\{\bs{\epsilon}_1^\prime,\bs{\epsilon}_2^\prime,\dots,$
$\bs{\epsilon}_N^\prime\}\in\E(\FS^N)$, we associate a basis $\{\bs{E}_0^\prime,
\bs{E}_1^\prime,\dots,\bs{E}_{N^2-1}^\prime\}$ in $\Herm(N)$ such that
$$
E^{\prime b\dot c}_\alpha\{\bs{\epsilon}_1^\prime,\dots,\bs{\epsilon}_N^\prime
\}=E^{b\dot c}_\alpha,\eqno(10)
$$
where $E^{b\dot c}_\alpha=E^{b\dot c}_\alpha\{\bs{\epsilon}_1,\dots,
\bs{\epsilon}_N\}$ and $\alpha=0,1,\dots,N^2-1$. In other words, (10) defines
the mapping $\{\bs{\epsilon}_1^\prime,\bs{\epsilon}_2^\prime,\dots,
\bs{\epsilon}_N^\prime\}\mapsto\{\bs{E}_0^\prime,\bs{E}_1^\prime,\dots,
\bs{E}_{N^2-1}^\prime\}$ of $\E(\FS^N)$ into the set of all bases in
$\Herm(N)$. However,
$$
E^{\prime b\dot c}_\alpha\{\bs{\epsilon}_1,\dots,\bs{\epsilon}_N\}=
c^b_f\overline{c^{\dot c}_{\dot g}}\,E^{\prime f\dot g}_\alpha
\{\bs{\epsilon}_1^\prime,\dots,\bs{\epsilon}_N^\prime\}\eqno(11)
$$
(compare it with (8)). Due to (10) and (11), we obtain
$$
E^{\prime b\dot c}_\alpha\{\bs{\epsilon}_1,\dots,\bs{\epsilon}_N\}=c^b_f
\overline{c^{\dot c}_{\dot g}}\,E^{f\dot g}_\alpha.\eqno(12)
$$

Let us consider the following expansions
$$
\bs{E}_\alpha^\prime=L(\text{C}_N)_\alpha^\beta\bs{E}_\beta,\eqno(13)
$$
where $L(\text{C}_N)_\alpha^\beta\in\R$ and $\alpha,\beta=0,1,\dots,
N^2-1$. In order to find $L(\text{C}_N)_\alpha^\beta$ as the functions of
$c^a_b$, it is useful to introduce $N^2$ Finslerian $N$-spintensors
$\bs{E}^\alpha$ of the valency $\left[{0\atop 1}{0\atop 1}\right]$ such that
$$
\contr(\bs{E}^\alpha\otimes\bs{E}_\beta)=\delta^\alpha_\beta.
\eqno(14)
$$
It is easy to show that $\bs{E}^\alpha$ exist, are unique, and $E^\alpha_{b\dot
c}=\overline{E^\alpha_{c\dot b}}$ with the notation $E^\alpha_{b\dot c}=
E^\alpha_{b\dot c}\{\bs{\epsilon}_1,\dots,\bs{\epsilon}_N\}$. Using (13) and
(14), we can write
$$
L(\text{C}_N)^\alpha_\beta=\contr(\bs{E}^\alpha\otimes\bs{E}_\beta^\prime).
\eqno(15)
$$
On the other hand, (12) implies
$$
\contr(\bs{E}^\alpha\otimes\bs{E}_\beta^\prime)=E^\alpha_{b\dot c}
c^b_f\overline{c^{\dot c}_{\dot g}}E^{f\dot g}_\beta.\eqno(16)
$$
Thus, according to (15) and (16),
$$
L(\text{C}_N)^\alpha_\beta=E^\alpha_{b\dot c}c^b_f\overline{c^{\dot c}_{\dot
g}}E^{f\dot g}_\beta.\eqno(17)
$$

Let $\text{E}^\alpha=\|E^\alpha_{\dot cb}\|$, $\text{E}_\beta=\|E_\beta^{f\dot
g}\|$, and $\text{E}_\beta^\prime=\|E_\beta^{\prime f\dot g}\{\bs{\epsilon}_1,
\dots,\bs{\epsilon}_N\}\|$. Then, it is possible to rewrite (12) and (17) in
the matrix form respectively as
$$
\text{E}_\beta^\prime=\text{C}_N\text{E}_\beta\text{C}_N^+\eqno(18)
$$
and
$$
L(\text{C}_N)^\alpha_\beta=\tr(\text{E}^\alpha\text{C}_N\text{E}_\beta
\text{C}_N^+),\eqno(19)
$$
where the cross denotes Hermitian conjugating. However, it follows from
(13) that $\text{E}_\beta^\prime=L(\text{C}_N)^\gamma_\beta\text{E}_\gamma$.
Therefore,
$$
\text{C}_N\text{E}_\beta\text{C}_N^+=L(\text{C}_N)^\gamma_\beta\text{E}_\gamma.
\eqno(20)
$$ 
Taking into account (19) and (20), we immediately obtain
$$
L(\text{B}_N\text{C}_N)^\alpha_\beta=\tr(\text{E}^\alpha\text{B}_N\text{C}_N
\text{E}_\beta\text{C}_N^+\text{B}_N^+)=L(\text{B}_N)^\alpha_\gamma
L(\text{C}_N)^\gamma_\beta\eqno(21)
$$
for any $\text{B}_N,\text{C}_N\in\SL(N,\C)$.

Let $L(\text{C}_N)=\|L(\text{C}_N)^\alpha_\beta\|$ and $\text{FL}(N^2,\R)=
\{L(\text{C}_N)\mid\text{C}_N\in\SL(N,\C)\}$. In these terms, (21) means that
$\text{FL}(N^2,\R)$ is a group with respect to the matrix multiplication and
the mapping
$$
L\colon\SL(N,\C)\to\text{FL}(N^2,\R),\quad\text{C}_N\mapsto L(\text{C}_N)
\eqno(22)
$$
is a group epimorphism so that, in particular, $L(1_N)=1_{N^2}$ ($1_N$,
$1_{N^2}$ are the identity matrices of the corresponding orders) and
$L(\text{C}_N^{-1})=L(\text{C}_N)^{-1}$. It is easy to prove that the kernel of
the epimorphism (22) has the form
$$
\ker L=\{e^{i\frac{2\pi k}{N}}1_N\mid k=0,1,\dots,N-1\}.\eqno(23)
$$

Let us return to the relations (13). Since both $\{\bs{E}_0,\dots,
\bs{E}_{N^2-1}\}$ and $\{\bs{E}_0^\prime,\dots,$ $\bs{E}_{N^2-1}^\prime\}$ are
bases in $\Herm(N)$, any vector $\bs{X}\in\Herm(N)$ can be expanded in the two
ways
$$
\bs{X}=X^\alpha\bs{E}_\alpha=X^{\prime\beta}\bs{E}^\prime_\beta,\eqno(24)
$$
where $X^\alpha$, $X^{\prime\beta}\in\R$. It is obvious that $X^{\prime\beta}=
L(\text{C}_N^{-1})^\beta_\alpha X^\alpha$. On the other hand, $X^{b\dot c}\{
\bs{\epsilon}_1,\dots,\bs{\epsilon}_N\}=c^b_f\overline{c^{\dot c}_{\dot g}}\,
X^{f\dot g}\{\bs{\epsilon}_1^\prime,\dots,\bs{\epsilon}_N^\prime\}$ or, what is
the same,
$$
\|X^{b\dot c}\{\bs{\epsilon}_1,\dots,\bs{\epsilon}_N\}\|=\text{C}_N\|X^{f\dot
g}\{\bs{\epsilon}_1^\prime,\dots,\bs{\epsilon}_N^\prime\}\|\text{C}_N^+.
\eqno(25)
$$
Remembering that $\det\text{C}_N=1$ and calculating the determinant of (25), we
see that
$$
\det\|X^{b\dot c}\{\bs{\epsilon}_1,\dots,\bs{\epsilon}_N\}\|=\det\|X^{f\dot g}
\{\bs{\epsilon}_1^\prime,\dots,\bs{\epsilon}_N^\prime\}\|\eqno(26)
$$
for {\it any\/} $\{\bs{\epsilon}_1^\prime,\dots,\bs{\epsilon}_N^\prime\}\in
\E(\FS^N)$. Hence, (26) gives an invariant numerical characteristic of the
vector $\bs X$, which is naturally denoted by $\det\bs{X}$. Notice that $\det
\bs{X}=\det\|X^{b\dot c}\{\bs{\epsilon}_1,\dots,\bs{\epsilon}_N\}\|\in\R$ as
it follows from (9).

Thus, without loss of generality, it is possible to calculate $\det\bs{X}$ with
respect to the basis $\{\bs{\epsilon}_1,\dots,\bs{\epsilon}_N\}\in\E(\FS^N)$.
According to (24) and (26), $\det{\bs X}=\det(X^\alpha\text{E}_\alpha)=
\det(X^{\prime\beta}\text{E}^\prime_\beta)$. However, (18) implies
$\det(X^{\prime\beta}\text{E}^\prime_\beta)=\det(\text{C}_N X^{\prime\beta}
\text{E}_\beta\text{C}_N^+)$ $=\det(X^{\prime\beta}\text{E}_\beta)$. Therefore,
$$
\det{\bs X}=\det(X^\alpha\text{E}_\alpha)=\det(X^{\prime\alpha}
\text{E}_\alpha).\eqno(27)
$$
At the same time,
$$
\det(X^\alpha\text{E}_\alpha)=G_{\alpha\beta\dots\gamma}\underbrace{X^\alpha
X^\beta\cdots X^\gamma}_{\text{$N$ multiplicands}},\eqno(28)
$$
where the real coefficients $G_{\alpha\beta\dots\gamma}$ are completely
determined by the choice of the basis $\{\bs{E}_0,\dots,\bs{E}_{N^2-1}\}$ in
$\Herm(N)$. Because of (27) and (28),
$$
\det\bs{X}=G_{\alpha\beta\dots\gamma}X^\alpha X^\beta\cdots X^\gamma=
G_{\alpha\beta\dots\gamma}X^{\prime\alpha}X^{\prime\beta}\cdots
X^{\prime\gamma},\eqno(29)
$$
i.e., $\det\bs{X}$ is {\it forminvariant\/} under transformations of the group
$\text{FL}(N^2,\R)$. Notice that (29) is valid for {\it any\/} basis
$\{\bs{E}_0^\prime,\dots,\bs{E}_{N^2-1}^\prime\}$ whose elements are connected
with those of $\{\bs{E}_0,\dots,\bs{E}_{N^2-1}\}$ by the relations (13).

Denoting $\det\bs{X}$ by $\bs{X}^N$ and using (28), we get (with respect to the
basis $\{\bs{E}_0,\dots,\bs{E}_{N^2-1}\}$)
$$
\bs{X}^N=G_{\alpha\beta\dots\gamma}X^\alpha X^\beta\cdots X^\gamma,\eqno(30)
$$
where $G_{\alpha\beta\dots\gamma}$ are symmetric in all the indices and do not
depend on the choice of any canonical basis in $\FS^N$. Thus, (30) correctly
defines the structure of an {\it $N^2$-dimensional flat Finslerian space\/} on
$\Herm(N)$ so that $\bs{X}^N$ is the $N$-th power of the Finslerian length of
the vector $\bs{X}\in\Herm(N)$~\cite{8}. It should be noted that, in general,
the homogeneous algebraic form (30) is not positive definite.

\section*{Conclusion}

In the present report, we have considered algebraic aspects of the Finslerian
$N$-spinor theory. We formulated the general definitions of a Finslerian
$N$-spinor and Finslerian $N$-spintensor of an arbitrary valency. It was shown
that Finslerian $N$-spintensors of the valency $\left[{1\atop 0}{1\atop 0}
\right]$ were closely associated with the $N^2$-dimensional flat Finslerian
space $\Herm(N)$. The metric on $\Herm(N)$ was characterized by the homogeneous
algebraic form (30) of the $N$-th power. We also constructed the generalization
(22) of the well known epimorphism $\SL(2,\C)\to\text{O}_+^\uparrow(1,3)$ and
found that its kernel consisted of the $N$ scalar matrices (23). In particular,
Finslerian 2-spinors coincide with standard Weyl 2-spinors. Indeed, for $N=2$,
the functional (1) is the ordinary symplectic scalar multiplication on $\FS^2$,
while $\Herm(2)$ is isomorphic to the 4-dimensional Minkowski space (we assume
$\text{E}^\alpha=\frac{1}{2}\sigma^\alpha$, $\text{E}_\alpha=\sigma_\alpha$,
where $\alpha=0$, 1, 2, 3 and $\sigma^\alpha=\sigma_\alpha$ are the identity
and Pauli matrices).

\end{document}